\documentclass[preprint,showpacs,titlepage,aps,prd,tightenlines,amsmath,byrevtex,nofootinbib]{revtex4}

\usepackage{graphicx}

\newcommand{\be}{\begin{equation}}
\newcommand{\ee}{\end{equation}}
\def\aprle{\buildrel < \over {_{\sim}}}
\def\aprge{\buildrel > \over {_{\sim}}}
\newcommand{\bea}{\begin{eqnarray}}
\newcommand{\eea}{\end{eqnarray}}

\begin{document}

\title{Neutrino Velocity and Neutrino Oscillations}

\author{H.~Minakata$^{1}$}
\email{hisakazu.minakata@gmail.com}
\author{A.~Yu.~Smirnov$^{2}$}
\email{smirnov@ictp.it}

\affiliation{
$^1$Department of Physics, Tokyo Metropolitan University, Hachioji, Tokyo 192-0397, Japan \\
$^2$The Abdus Salam International Centre for Theoretical Physics,\\
Strada Costiera 11, I-34013 Trieste, Italy 
}

\vglue 1.2cm

\begin{abstract}

We study  distances of propagation and the group velocities of the muon neutrinos 
in the presence of mixing and oscillations assuming that Lorentz invariance holds.  
Oscillations lead to distortion of  the $\nu_\mu$ wave packet which, in turn, changes 
the group velocity and the distance $\nu_\mu$ travels. 
We find that the change of the distance, $d_{osc}$, is proportional to the length of the wave packet, $\sigma_x$, and the oscillation phase, $\phi_p$, acquired by neutrinos in the  $\pi-$ and  $K-$ meson decay tunnel where neutrino wave packet is formed: $d_{osc} \propto \sigma \phi_p$.  
Although the distance $d_{osc}$ may effectively correspond to the superluminal motion,  the effect is too tiny ($\sim 10^{- 5}$ cm ) to be reconciled with the OPERA result.  We analyze various possibilities to increase $d_{osc}$ and discuss experimental setups in which  $d_{osc}$ (corresponding to the superluminal motion) can reach an observable value $\sim 1$ m.

\end{abstract}

\pacs{03.30.+p,14.60.Pq,14.60.Lm}

\date{June 13, 2012}

\maketitle

\section{Introduction}

Neutrinos are the lightest known massive particles and therefore their velocity at accessible energies can be the closest one to the velocity of light. 
The velocity of neutrinos (and therefore the difference of the velocities of neutrinos and photons) can be affected by various factors: 
gravitational field, modifications of metric, and local space-time environment, matter through which neutrinos propagate, local effects induced by presence of new fields in space, possible interactions with ``dark'' components in the Universe, etc. Last but not least the velocity can be affected by  Lorentz violation. Therefore, measurements of neutrino velocity may reveal new phenomena related to physics beyond the Standard Model. They provide important 
probe of properties of space and time as well as dark sector of the Universe.   

At high energies neutrino velocity has been measured in several accelerator experiments \cite{Fermilab,MINOS,OPERA,ICARUS}. At low energies, observations of neutrino burst from SN1987a \cite{SN1987a} place severe constraints on neutrino velocity. Many of the factors mentioned above, which affect neutrino velocity,  have been discussed in detail \cite{many} in connection to the controversial OPERA result \cite{OPERA}. 

To address properly the issue of possible superluminal velocity of particles one must be aware of that it can arise even within the conventional framework based on special theory of relativity. In fact, superluminal propagation of light is a well known subject in optics \cite{Garrett}. The effect is related to distortion of the photon pulse during propagation in media, so that suppression of trailing edge of the pulse leads to increase of the group velocity. The superluminal motion does not contradict causality and no information can be transmitted with velocity larger than the velocity of light. Indeed, the effect has been observed in several experiments (see e.g., \cite{Chu,Wang,Sanchez,Brunner}). 

A similar mechanism for neutrinos has been discussed in a series of papers \cite{Mecozzi,Morris,Berry,Indumathi,Tanimura} in the context of analyzing the OPERA result. 
In \cite{Mecozzi,Morris,Berry,Indumathi} 
the superluminal motion of the muon neutrinos has been considered
in the presence of flavor mixing, in which muon neutrino is described
by coherent combination of the mass eigenstates.
Distortion of the $\nu_\mu$ wave packet  is produced by interplay of two effects:
(i) the coordinate independent $\nu_\mu$ oscillations in time,
and (ii) a relative shift of the wave packets of mass eigenstates
due to difference of their group 
velocities.\footnote{
We will give detailed explanation of the effect in the Appendix.
}
In this proposal an additional contribution to the distance of 
$\nu_\mu$ propagation, $d$, and to the group velocity of muon neutrinos is proportional to the difference of
the group velocities of the neutrino mass eigenstates $\Delta v \approx  \Delta m^2/2E^2$, which is extremely small. Since  $d \sim 1/P_{\mu\mu}$, 
the effect can be enhanced  if the survival probability
of muon neutrino, $P_{\mu\mu}$,  is very small.  
The latter, in turn, requires mixing to be very close to the maximal and
the oscillation phase to be very close to $\pi$. Apparently an additional distance of $\nu_\mu$ propagation is restricted by the length of neutrino wave packet.

In this paper, we give a comprehensive treatment of the group velocities of neutrinos in vacuum and in matter within the framework of Lorentz invariance. 
As in the earlier works \cite{Mecozzi,Morris,Berry,Indumathi,Tanimura} we consider the superluminal motion of the muon neutrinos in the presence of flavor mixing, in which muon neutrino is described by coherent superposition of the neutrino mass eigenstates. 
We show that the dominant effect, which most affects the velocity of muon neutrinos, is a distortion of the $\nu_\mu$ wave packet by the neutrino oscillations within the size of $\nu_\mu$ wave packet in the coordinate space. The distortion  can be significant because neutrino wave packets are large, as will be discussed in Sec.~\ref{sec:pdecay}. On the other hand, the effect of relative shift of the wave packets of mass eigenstates due to their velocity difference is negligible compared to the dominant one. 

The paper is organized as follows. In Sec. II we  
present general formulas for the distance traveled by oscillating 
neutrinos during a given interval of time. 
In Sec. III we construct a wave packet of the muon 
neutrino produced in the pion decay  and describe its properties. 
In Sec. IV we give explicit expressions for the distances traveled by muon neutrinos. 
We consider the limit of small size of decay tunnel
when the wave packet has approximately the box-like (rectangular) shape. Effect of
relative shift of the wave packets of mass eigenstates on the distance is computed. In Sec. V we estimate the distances and velocities for existing experiments and propose experimental setups in which the effective 
superluminal motion might be observed.
In Sec.VI we consider group velocities of neutrinos in matter. 
We conclude in Sec. VII. In the Appendix 
a simple explanation of the shift effect proposed before is given.

\section{Oscillation probability and distance of $\nu_\mu$ propagation} 
\label{sec:probdist}

For simplicity we will consider the two neutrino mixing  
\be
\nu_\mu = c \nu_1 + s \nu_2, ~~~ \nu_\tau = c \nu_2 - s \nu_1, 
\label{eq:mixing}
\ee 
where $c \equiv \cos \theta$, $s \equiv \sin \theta$, and  $\theta$ is  the mixing 
angle;  $\nu_i$ $(i = 1, 2)$ are the mass eigenstates which correspond to eigenvalues $m_i$.  We will show in Sec.~\ref{sec:matter} that this is a good approximation for the three neutrino mixing case. 
Recall that in experiments such as OPERA, MINOS, T2K, {\it etc.}, 
the neutrinos oscillate in matter, and therefore in the presence of 
non-zero 1-3 mixing one should use the mixing angles and mass 
splitting  in matter. However,  apart from the resonance regions  
$E \sim 0.1$ GeV and $E = (4 - 8)$ GeV, the two neutrino 
description with vacuum values of the mixing angles and 
$\Delta m^2 \equiv m_2^2 - m_1^2$  
gives a very good approximation.   
 
Evolution of the muon neutrino state  
after it exits the region of formation of  wave packets,  
{\it i.e.} a decay tunnel, is described as 
\be
|\nu_{\mu}(t)\rangle  = c f_1 (x - v_1 t) e^{-i E_1 t + i p_1 x} 
|\nu_1 \rangle + s f_2 (x - v_2 t) e^{-i E_2 t + i p_2 x}  |\nu_2 \rangle,  
\label{eq:numu}
\ee
where $f_i$ are the shape factors, 
$E_i$ and $p_i$ are the average energies and momenta of the 
wave packets, respectively, and $v_i$  are the group velocities of the mass 
eigenstates.  The shape factors are normalized as 
\begin{eqnarray}
\int dx \vert f_i(x - v_i t) \vert^2 = 1.   
\label{normalization-def}
\end{eqnarray} 

If muon neutrino is detected at time $t$,  
the wave function of $\nu_\mu$ (the amplitude of probability to find $\nu_\mu$) is given by 
\be
\psi_{\nu_\mu}(x, t) = \langle \nu_\mu|\nu_{\mu}(t)\rangle = 
c^2 f_1 (x - v_1 t) e^{-i E_1 t + i p_1 x} + 
s^2 f_2 (x - v_2 t) e^{-i E_2 t + i p_2 x}.   
 \label{eq:psimu}
\ee
Here, we assumed that the shape factor 
of detected muon neutrino is constant in the 
detection area and does not depend on time. 
Alternatively, one can consider $f_i$ as the effective 
shape factors which already include process of detection. 

The $\nu_\mu$ survival probability at $t$ is obtained as 
\be
P_{\mu\mu}(t)  \equiv  \int dx~\vert \psi_{\nu_\mu} (x, t) \vert^2.
\label{eq:Pmumu-def}
\ee 
In what follows, we will compute the averaged coordinate $\langle x (t) \rangle$ 
of the muon neutrino at time $t$ defined as 
\begin{eqnarray}
\langle x (t) \rangle \equiv  
\frac{ \int dx ~x ~\vert \psi_{\nu_\mu} (x, t) \vert^2 }{ 
\int dx~\vert \psi_{\nu_\mu} (x, t) \vert^2 }, 
\label{eq:avarage-x-def}
\end{eqnarray}
where the  denominator is nothing but $P_{\mu\mu}(t)$. 

In a very good approximation, we can take at the initial time 
\be
f_1(x) = f_2(x) 
\label{eq:equal}
\ee
so that the difference between $f_1$ and $f_2$ at an arbitrary time arises solely due to the difference of group velocities of  $\nu_1$ and $\nu_2$. 
The difference $\Delta v \equiv v_1 - v_2$ produces a relative shift and eventually separation of the wave packets of mass eigenstates in the process of propagation. 
The separation leads to loss of coherence between the mass eigenstates. 
Numerically, 
$$
\Delta v \equiv v_1 - v_2 \simeq \frac{ \Delta m^2 }{ 2 E^2} 
= 5 \times 10^{-22} \left( \frac{ \Delta m^2 }{ 10^{-3} \text{eV}^2} \right) 
\left( \frac{ E }{ 1 \text{GeV} } \right)^{- 2}.  
$$

Using the expression (\ref{eq:psimu}) and the normalization condition 
(\ref{normalization-def}) we find 
\be 
P_{\mu\mu}(t)  = 
c^4 + s^4 + 2 c^2 s^2 I , 
\label{eq:Pmumu1}
\ee
where 
\be
I \equiv \int dx ~f_1(x - v_1 t) f_2(x - v_2 t) \cos \Phi(x, t),  
\label{eq:I-def}
\ee
and the relative (oscillation) phase 
of mass eigenstates, $\Phi(x, t)$,  is given by 
$$
\Phi(x, t) \equiv  \Delta E t - \Delta p x, ~~~\Delta E \equiv E_1 - E_2,~~~ 
\Delta p \equiv p_1 - p_2. 
$$
The phase $\Phi(x, t)$ can be split into two pieces as 
\be
\Phi(x, t) \equiv  \phi  +  \chi,  
\label{eq:phase}
\ee
where 
\be
\phi \equiv \Delta E t - \Delta p vt, ~~~~~ v \equiv \frac{v_1 + v_2}{2} 
\label{eq:phi-t}
\ee
is the standard oscillation phase which depends 
on time only and 
\be
\chi \equiv - \Delta p (x - vt). 
\label{eq:phchi}
\ee
The phase $\chi$ depends also on the coordinate $x$ and,  
as we will see,  describes the change of the total oscillation 
phase within the wave packets.    

Let us introduce the average coordinates of the mass eigenstates as 
\be
\langle x_i \rangle \equiv  
\int dx ~x ~[f_i (x - v_i t)]^2 = v_i t  + 
\int dz~ z ~[f_i (z)]^2. 
\label{av-x-def}
\ee
If $f_i(z)$ is symmetric with respect to the  central point 
of packet, $z_0$: $f_i(z_0 - h) = f_i(z_0 + h)$, 
we obtain (performing shift of integration $z \rightarrow z - z_0$):   
$\langle x_i \rangle = v_i t + z_0$.  

Using  $\langle x_i \rangle$ we define the averaged coordinate 
of the two mass eigenstates, 
\be
\bar{x} \equiv 
\frac{1}{2}(\langle x_1 \rangle + \langle x_2 \rangle) 
= \frac{1}{2}(v_1 + v_2) t  
+ \int dz~ z ~[f_i (z)]^2,
\label{eq:x-av}
\ee
and the half-difference of the coordinates
$$
\Delta x \equiv \frac{1}{2}(\langle x_1 \rangle - \langle x_2\rangle) 
=  \frac{1}{2} \Delta v t.  
$$

Then, the numerator of $\langle x(t) \rangle$ in (\ref{eq:avarage-x-def}) 
can be written as 
\be 
\int dx ~x ~\vert \psi_{\nu_\mu} (x, t) \vert^2   
= (c^4 + s^4)\bar{x} + \frac{\Delta v t}{2} + 2 s^2 c^2 J,
\label{eq:numerator}
\ee
where 
\be
J \equiv  
\int dx ~x ~f_1(x - v_1 t) f_2(x - v_2 t) \cos \Phi(x, t) 
\label{eq:J-def}
\ee
is the average coordinate of the overlap of the two mass eigenstates.  

Using  definition (\ref{eq:avarage-x-def}),  (\ref{eq:Pmumu1}) 
and  (\ref{eq:numerator})  
we can present $\langle x(t) \rangle$ in  the form  
\be 
\langle x(t) \rangle = \bar{x} + 
\frac{\Delta v t \cos 2\theta}{2P_{\mu\mu}} + 
\frac{\sin^2 2\theta}{2P_{\mu\mu}}(J - \bar{x}I).
\label{eq:average-x1}
\ee
The first term in (\ref{eq:average-x1}) 
is the distance traveled by massive neutrinos,   
and the two others describe the additional distances due 
to oscillations and relative shift of the wave packets.  

The distance of $\nu_\mu$ propagation  
during a time interval from $t_0$ to $t$ equals  
\be
d(t) \equiv \langle x(t) \rangle - \langle x(t_0) \rangle. 
\label{eq:d-def}
\ee
As $t_0$  we will take the moment in time when the neutrino 
wave packets are completely formed. For  neutrinos from pion decay 
$t_0 = l_p/v_\pi$  is the moment of time when 
pion reaches the end of decay tunnel.  
According to (\ref{eq:average-x1})
the distance $d(t)$ can be presented as 
\be  
d(t, t_0) = d_{light} + d_{mass} + d_{osc-a} + d_{shift-s}, 
\label{eq:decomposition}
\ee
where   
$$
d_{light} \equiv c(t - t_0), 
$$
is the distance traveled by the light, 
$$
d_{mass} = - \frac{\bar{m}^2}{2E^2} (t - t_0) = 
- \frac{m_1^2 + m_2^2}{4E^2} (t - t_0) 
$$
(where $E$ is the neutrino energy) is the contribution 
to the distance due to non-zero neutrino mass; 
$$
d_{osc-a} = - \frac{\sin^2 2 \theta}{2}
\left[ \left. \frac{J - \bar{x} I}{P_{\mu \mu}}\right|_{t} - 
\left. \frac{J - \bar{x} I }{P_{\mu \mu}} \right|_{t_0} \right]
$$
is, as we will see, the contribution to the distance from oscillation distortion 
of the muon neutrino wave packet and from shift of the packets 
in the case of asymmetric shape factors; and   
\be
d_{shift-s} = \frac{\Delta v \cos 2\theta}{2} 
\left[\frac{t}{P_{\mu \mu} (t)} -  \frac{t_0}{P_{\mu \mu} (t_0)}
\right]
\label{eq:shift-s}
\ee
is the contribution from the relative shift of the wave packets and 
coordinate-independent oscillations in the case of symmetric shape factors.  
Notice that the terms in the brackets evaluated 
at $t=t_0$ are negligible, since usually $t \gg t_0$ and $P_{\mu \mu}(t_0) \sim 1$. 
However, it may not be always the case for $d_{osc-a}$ because 
$(J - \bar{x} I)$ is non-linear function of $t$. 

As we will see, the distance traveled by muon 
neutrinos is an oscillatory function of time. 
Correspondingly,  the velocity changes with time. Therefore,   
we can speak about the average velocity for a given 
time interval $(t - t_0)$ defined as  $v(t) = d(t, t_0)/(t - t_0)$.

\section{Neutrino wave packets from pion decay }
\label{sec:pdecay}
  
For definiteness we will consider neutrinos from pion decay. 
Actually, these neutrinos dominate in the neutrino fluxes of the MINOS, 
T2K and OPERA experiments. The contribution 
from $K-$ mesons can be considered similarly. 
We  describe here the wave packets of muon neutrinos (the wave function
$\psi_{\nu_\mu}$, in  the eq. (\ref{eq:psimu})).  
The strict derivation of the expressions for the wave packets 
is given in \cite{AHS}, and here we explain properties of  
the shape factors and phases using simple physics arguments.  

Let us consider first the shape factors. 
Pions are produced in the strong interaction of protons 
with a solid state target. 
Protons have wave packets of very small size and can be considered as point-like. 
The target nuclei are localized  within the atomic scale. 
Therefore we can take in our computations that pions are produced 
at fixed space-time point $x = t = 0$.  
The wave packets of pions are very short in the configuration space,  
so that pions also can be considered as point-like with space-time trajectories 
of motion $x = v_\pi t$. Here $v_\pi$ is the group velocity of 
pion.  In experiments under consideration pions are ultra-relativistic 
with Lorentz factor $\gamma_\pi \equiv E_\pi/m_\pi \gg 1$,  
where $E_\pi$ and $m_\pi$ are the pion energy and the mass, respectively. 

Pions decay in a decay tunnel of size $l_p$ which includes the length of the 
horn area and the decay pipe. 
Essentially $l_p$ is the distance from the target to the end point of decay 
pipe where survived pions are absorbed. We neglect the effects of 
interactions of pions with particles in the decay pipe. 
In the decay tunnel the quantum state is a superposition of un-decayed 
pion state and a state of muon and muon-neutrino. 
The neutrino waves emitted from different points of the pion 
trajectory are  coherent, 
thus forming a single wave packet.  
 
The decay length of pion equals 
$$
l_{decay} =  v_\pi \tau_\pi = v_\pi \gamma_\pi \tau_\pi^0,  
$$  
where $\tau_\pi$ and $\tau_\pi^0$ are the lifetime of pion in the laboratory frame 
and the one in the pion rest frame, respectively. 
$\tau_\pi^0 = \Gamma_{0}^{-1}$ with $\Gamma_{0}$ being the pion decay 
rate at rest.

If $l_p \leq l_{decay}$ (high energies, 
long-lived parent particles, short decay tunnel), 
the wave packets of mass eigenstate $\nu_i$ have the size 
\be
\sigma_i = l_p \left(\frac{v_i}{v_\pi} - 1\right) \approx \frac{l_p}{ 2 \gamma_\pi^2}, 
\label{eq:wpsize}
\ee
when neutrino is emitted in the forward direction with respect to the pion 
velocity. The wave packet size is proportional to $l_p$;   
the factor in bracket represent shrinking of the packet due to pion motion~\cite{farzan}. 
In the second equality 
in (\ref{eq:wpsize}) we used  $m_i \ll m_\pi$.   
Numerically~\footnote{
For two body pion decay the neutrino energy $E$ is determined by Lorentz 
factor of pion, $\gamma_\pi$, or {\it vice versa}: 
$\gamma_\pi \simeq 16.8  \left( \frac{ E }{ 1 \text{GeV} } \right)$ to a good approximation 
for $E \geq 500$ MeV and forward going neutrinos.
}
\be
\sigma  
 \simeq 50 \text{ cm }
 \left( \frac{ \gamma_\pi }{ 10} \right)^{- 2} 
 \left( \frac{ l_p }{ 100 \text{m} } \right) 
=  17.7 \text{ cm }
\left( \frac{ E }{ 1 \text{GeV} } \right)^{- 2} 
 \left( \frac{ l_p }{ 100 \text{m} } \right).    
 \label{eq:packet-size}
\ee
%

If $l_{decay} \ll l_p$ 
(low energies, long decay tunnel, short-lived parent particles),  
the size of the wave packet is determined by the decay length:
\be
\sigma_i = l_{decay} \left(\frac{v_i}{v_\pi} - 1 \right)
=  \gamma_\pi \tau_\pi^0 (v_i - v_\pi).
\label{eq:sigmadec}
\ee
For relativistic pions ($\gamma_\pi \gg 1$), 
$\sigma \approx  \frac{1}{2} \tau_\pi^0 \gamma_\pi^{-1}$ 
instead of (\ref{eq:wpsize}).  In the non-relativistic limit, $v_\pi \rightarrow 0$, 
the  size becomes the largest: $\sigma_i = \tau_\pi^0  v_i \simeq 7.8$ m. 

Thus, in general, 
\be
\sigma_i = l_{form} \left(\frac{v_i}{v_\pi} - 1\right) \approx \frac{l_{form}}{ 2 \gamma_\pi^2}, 
\label{eq:wpsize1}
\ee
where $l_{form}$ is the region of formation of the neutrino wave packet:
$$
l_{form} \sim {\rm min} \left\{ {l_p, l_{decay}} \right\} .
$$

In what follows we will neglect difference between 
the sizes of wave packets 
of the  mass eigenstates, taking $\sigma_1 \approx \sigma_2$. 
It is an excellent approximation because the difference is of the order 
$\sim l_p \Delta v \sim 10^{-16}$cm for $l_p = 1$ km,  $E = 1$ GeV 
and $\Delta m^2 = 2.5 \cdot 10^{-3}$ eV$^2$. 

The neutrino shape factor from pion decay can be written 
as~\cite{AHS} 
\be
f (x) = f_0 e^{\frac{\Gamma}{2(v -  v_\pi)} (x - \sigma)} 
\Pi \left(x, [0, \sigma] \right), 
\label{expprof}
\ee
where $\Pi \left(x, [0, \sigma] \right) = 1$ in the interval 
$0 \leq x \leq \sigma$ and $\Pi$ vanishes outside this interval. 
The box function reflects a finite size of neutrino production region  
with sharp edges related to the definite points of pion production 
and absorption at the end of decay tunnel. The exponential factor  
follows from the amplitude of the pion decay with the rate 
enhanced by the factor $1 / (v - v_\pi)$ which corresponds to 
shrinking of the neutrino wave packet emitted in forward direction  
(see eqs. (\ref{eq:wpsize1})). This is nothing but the Doppler effect. 

The normalization condition (\ref{normalization-def}) gives 
\begin{eqnarray}
f_0^{2} =  \frac{y}{\sigma} \frac{1}{ 1 - e^{-y} }~ , 
\label{normalization}
\end{eqnarray}
where 
\be
y  \equiv \frac{\Gamma \sigma }{v - v_\pi} = \frac{l_{form}}{l_{decay}}.   
\label{y-def}
\ee
If $l_{decay} >  l_p$, we have $l_{form} = l_p$, so that   
$y$ is the size of the decay tunnel in units of the decay length. 
Numerically, 
\bea
y &=& \frac{l_p}{l_{decay}} \approx \frac{ l_p \Gamma_0 }{ \gamma_\pi} \simeq 
1.28 \times \left( \frac{ \gamma_\pi }{ 10 } \right)^{- 1} 
\left( \frac{ l_p }{ 100 \text{m} } \right). 
\eea
If $l_{decay} <  l_p$ and if we neglect 
effects of the exponentially suppressed tails, then $l_{form} \simeq l_{decay}$, so that 
$y \simeq 1$. 

In the approximation $v_i \approx v$, the difference  
of momenta of the neutrino mass eigenstates equals~\cite{AHS}     
\be
\Delta p \equiv p_1 - p_2 
\approx - \frac{\Delta m^2}{2E(v - v_\pi)} \approx  
- \frac{\Delta m^2}{2E}2\gamma_\pi^2.  
\label{Delta-p}
\ee
The factor in the denominator of (\ref{Delta-p}) is the same as the one which describes 
shrinking of the wave packet size in comparison with size 
of the wave packet formation region. The shrinking 
is accompanied by increase of frequencies by the same factor. 
This is again the Doppler effect related to neutrino emission from moving pion.  
Consequently,  the oscillation effect  
within the packet is enhanced: 
It is determined by the oscillations within the wave packet formation region,  
rather than within the size of the packet itself in the laboratory frame.

\section{Distances of $\nu_\mu$ propagation}
\label{sec:v-oscillation}

We will use the  formulas derived in Sec. \ref{sec:probdist} and the 
wave function of the muon neutrino constructed in Sec. \ref{sec:pdecay} 
to find $P_{\mu \mu}$ and $\langle x(t) \rangle_{osc}$. 
We first compute the effect of oscillations, which gives the dominant effect, 
neglecting a relative shift of the $\nu_1$ and $\nu_2$ wave packets due to 
difference of group velocities. 
Then, we will estimate  the effect of the shift neglecting  
oscillations along the wave packets. 

\subsection{Oscillation effect}

Let us set $v_1 = v_2 = v$ ($\Delta v= 0$)  
and  compute $I$ and $J$ given in (\ref{eq:I-def}) 
and (\ref{eq:J-def}). We use the shape factors (\ref{expprof}) and  
the phase $\Phi$  with  $\phi$ defined in  (\ref{eq:phi-t})  
and $\chi$ given according to (\ref{eq:phchi}) and  (\ref{Delta-p}) by  
$$
\chi = \frac{\Delta m^2}{2E} \frac{x - {v} t}{v - v_\pi}.   
$$
We obtain the probability  
\be
P_{\mu \mu} (t) = c^4 + s^4 + 2 s^2 c^2 
(\cos \phi I_c - \sin \phi I_s ),  
\label{eq:mmprosc}
\ee
where 
\bea
I_c & \equiv & \int dz f^2(z) \cos |\Delta p| z 
  = \frac{y}{1 - e^{-y}} \frac{1}{y^2 + \phi_p^2} 
\left(- y e^{-y} + y \cos\phi_p + \phi_p \sin \phi_p \right), 
\nonumber\\  
I_s & \equiv & \int dz f^2(z) \sin |\Delta p| z 
= \frac{y}{1 - e^{-y}} \frac{1}{y^2 + \phi_p^2} 
\left(\phi_p e^{-y}  + y \sin \phi_p  -\phi_p \cos \phi_p \right).
\eea
Here  
\be
\phi_p \equiv \frac{\Delta m^2}{2E} \frac{\sigma}{(v - v_\pi)} \approx 
\frac{\Delta m^2}{2E} l_{form} = 2\pi \frac{l_{form}}{l_\nu}, 
\label{eq:phi-p-def}
\ee
is the phase change within the wave packet. Here $l_\nu = 4\pi E /\Delta m^2$ 
is the oscillation length. 
The second and third equalities in eq. (\ref{eq:phi-p-def}) are valid for an ultrarelativistic pion: 
$v_\pi \approx 1$. 
Thus, $\phi_p$ is given by 
the phase acquired within the distance of formation of the neutrino 
wave packet, $l_{form}$. For $l_p < l_{decay}$  we have 
\be
\phi_p = 
\frac{\Delta m^2}{2E} l_p \approx
2.5 \times 10^{- 4} \left( \frac{ \Delta m^2 }{ 10^{-3} \text{eV}^2} \right) 
\left( \frac{ E }{ 1 \text{GeV} } \right)^{- 1} 
\left( \frac{ l_p }{ 100 \text{m} } \right). 
\label{eq:phi-p-value}
\ee
For $l_p > l_{decay}$ inserting $\sigma$ from (\ref{eq:sigmadec}) 
into (\ref{eq:phi-p-def}) we obtain $\phi_p =  2\pi \gamma_\pi \tau^0_\pi v/l_\nu$. 

We now calculate $\langle x (t) \rangle$ (\ref{eq:average-x1}) 
for $\Delta v = 0$. 
The average distance traveled by mass eigenstates (\ref{eq:x-av})
with the wave packets (\ref{expprof}) equals  
\be 
\bar{x} = vt + \sigma \left(\frac{1}{1 - e^{-y}} - \frac{1}{y} \right). 
\label{eq:bar-x}
\ee
A straightforward computation of the contribution from the oscillation effect 
to the distance of $\nu_\mu$ propagation gives 
\be
\langle x \rangle_{osc} =  - \frac{\sin^2 2 \theta}{2 P_{\mu \mu}} 
\frac{\sigma}{y^2 + \phi_p^2} \left(\kappa_s \sin \phi - 
\kappa_c \cos \phi \right).   
\label{eq:osc-corr}
\ee
Here, the coefficients $\kappa_s$ and $\kappa_c$ in front of $\sin \phi$ and 
$\cos \phi$ equal 
\bea
\kappa_s & = & \frac{y}{1 - e^{-y}} \left[ 
\frac{-2 y\phi_p e^{-y}  + q \sin \phi_p - n\cos\phi_p}{ y^2 + \phi_p^2} \right. 
\nonumber\\ 
 && \hspace*{16mm} {} - \left. \left(\frac{1}{1 - e^{-y}} - \frac{1}{y} \right)
 \left(\phi_p e^{-y} + y \sin\phi_p - \phi_p \cos \phi_p \right) \right], 
\label{eq:rs}
\eea
\bea
\kappa_c & = & \frac{y}{1 - e^{-y}} \left[ 
\frac{(y^2 - \phi_p^2) e^{-y}  + q \cos \phi_p + n\sin\phi_p}{ y^2 + \phi_p^2} \right. 
\nonumber\\ 
 && \hspace*{16mm} {} - 
 \left. \left(\frac{1}{1 - e^{-y}} - \frac{1}{y} \right)
 \left(-y e^{-y} + y \cos \phi_p + \phi_p \sin \phi_p \right) \right], 
\label{eq:rc}
\eea
where  
$$
q  \equiv  y^2(y - 1) + \phi_p^2(y + 1), ~~~~~
n  \equiv  \phi_p (y^2 - 2 y + \phi_p^2).
$$

In realistic setups: $y \sim 1$ and $\phi_p \ll y $. 
For $\phi_p \rightarrow 0$  we obtain 
$\langle x \rangle_{osc} = 0$, because there is no room in which oscillation 
effect develops if $l_p \ll l_\nu$.  
In the limit of small  $\phi_p$ the distance equals 
\be
\langle x \rangle_{osc} =  - \frac{\sin^2 2 \theta}{2 P_{\mu \mu}} 
~\frac{\sigma \phi_p}{y^2} 
\left[ 
\eta_s \sin \phi - \eta_c  \frac{\phi_p}{y} \cos \phi 
\right], 
\label{eq:expeff}
\ee
where 
$$ 
\eta_s = 1 - \left(\frac{y}{1 - e^{-y}} \right)^2 e^{-y} , 
$$
$$ 
\eta_c = 2 - \left(\frac{y}{1 - e^{-y}} \right)\left(1 - y + \frac{y^2}{2} \right) -  
\left(\frac{y}{1 - e^{-y}} \right)^2 \left(1 - \frac{y}{2} \right) . 
$$
At $y = 1$, $\eta_s/y^2 \approx 1/12.61$, and it depends very weakly on $y$.   
It exactly equals to 1/12 in the case of box-like packet 
which corresponds to $y \rightarrow 0$ (see Sect. IVB).  

\begin{figure}[h]
\includegraphics[scale=0.80]{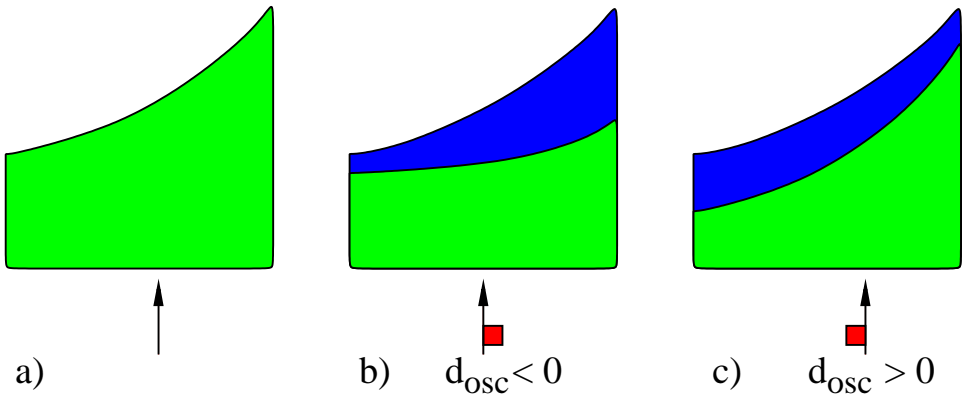}
\caption{The wave packets of the muon neutrino from pion decay 
without oscillations  (a) and with oscillations in two different moments of time (b),(c). The light-shadowed (green) parts of the shape factors show the $\nu_\mu$ fraction, 
whereas the dark-shadowed (blue) parts correspond to the $\nu_\tau$ fraction 
which appears due to oscillations. The arrows indicate  positions of 
``centers of mass'' of the $\nu_\mu$ parts. The small (red) boxes show 
shifts of the centers due to oscillations with respect to the center in the  
no-oscillation case.  The pannel (b)  
corresponds to the baselines $0 < L < l_{osc}/2$, when the front edge of the 
wave packet is suppressed. The pannel (c) is for $l_{osc}/2 < L < l_{osc}$, when 
the trailing edge is suppressed.}
\label{fig:pack}
\end{figure}

The effect of oscillations on the group velocity 
of the  muon neutrino is illustrated in the Fig.~\ref{fig:pack}.  
In the case of Fig.~\ref{fig:pack} (b), the oscillations suppress the front edge 
of the $\nu_\mu$ packet  stronger than the trailing edge. 
As a result, the ``center of mass'' of the packet shifts backward 
and therefore the distance of propagation is reduced. 
In the case of Fig.~\ref{fig:pack} (c) realized in another moment of time 
the trailing edge is suppressed more strongly; the center of mass shifts 
forward and the distance of propagation is increased.

\subsection{Limit of small production region, $y \ll 1$}

Expression for $\langle x \rangle$ obtained in Sec. IVA  simplifies 
substantially in the limit $y \rightarrow 0$. 
It corresponds to short decay tunnels $l_p \ll l_{dec}$,  
or to high energy parts of the neutrino spectra in real experiments.  
In this limit $\Gamma \rightarrow 0$ and therefore 
\be
f (x) = \frac{1}{\sqrt{\sigma}} \Pi \left(x, [0, \sigma] \right),  
\label{boxprof1}
\ee
and we will refer to this as the ``box-like'' packet. 

To obtain $P_{\mu\mu}$ and $\langle x \rangle_{osc}$ 
for the wave packet (\ref{boxprof1})
we take the limit $y \rightarrow 0$ in (\ref{eq:mmprosc}) and (\ref{eq:osc-corr}) 
while keeping $\phi_p$ finite.  The $\nu_{\mu}$ survival probability becomes  
\be
P_{\mu\mu}(t)    = 
c^4 + s^4 + \sin^2 2\theta 
\left[ \frac{ \sin \frac{\phi_p}{2} }{ \phi_p } \right]
\cos \left( \phi + \frac{\phi_p}{2} \right).  
\label{eq:Pmumu2}
\ee

In the limit $y \rightarrow 0$ 
the second term in (\ref{eq:bar-x}) tends to $\sigma/2$,  
and consequently,  $\bar{x} = v t + \frac{\sigma}{2}$. 

Finally, $\langle x \rangle_{osc}$ takes the form 
\be
\langle x (t) \rangle_{osc} =  
- \frac{\sin^2 2\theta}{2 P_{\mu\mu}} \frac{\sigma}{\phi_p^2} 
\left( 2 \sin \frac{\phi_p}{2} - \phi_p \cos \frac{ \phi_p }{2 } \right) 
\sin \left( \phi + \frac{\phi_p}{2} \right).  
\label{eq:average-x2}
\ee  
If $\phi_p$ is small (for existing setups $\phi_p \sim 10^{-3}$)
$\langle x \rangle_{osc}$ varies with distance traveled  as $\sim \sin \phi$.   
For short baselines, when $\phi < \pi/2$, the additional distance 
$\langle x (t) \rangle_{osc} < 0$, {\it  i.e.} oscillations suppress 
velocity (see Fig.~\ref{fig:pack}).  In addition, 
for small $\phi$ the distance is suppressed by $\phi$.    
At around $\pi/2$, $3\pi/2$, {\it etc.} the oscillation effect is maximal for $\langle x (t) \rangle_{osc}$. 
Furthermore, at $\phi \sim \pi/2$ ($3\pi/2$) the oscillation probability 
is a decreasing (increasing) function of $x$. 
Therefore, the effective shape of the $\nu_\mu$ wave packet 
deforms in such a way that the effective 
velocity is smaller (larger) than the normal velocity of massive particles at  
$\phi \sim \pi/2$ ($3\pi/2$).

In the limit  $\phi_p \rightarrow 0$, the  
expressions in (\ref{eq:Pmumu2}) and (\ref{eq:average-x2}) 
give 
\bea
P_{\mu\mu}(t) &\approx&  
c^4 + s^4 + \frac{1}{2} \sin^2 2\theta 
\left[ \cos \left(\phi + \frac{\phi_p}{2} \right)  -  \frac{\phi_p^2}{6}\cos \phi \right] ,  
\label{eq:Pmumu3}
\\
\langle x (t) \rangle_{osc} &\approx& 
- \frac{\sin^2 2\theta}{24 P_{\mu\mu}} 
\sigma \phi_p
\sin \left( \phi + \frac{\phi_p}{2} \right),   
\label{eq:average-x3}
\eea
where we have used the fact that the factor inside the parenthesis in front of the sine factor in  
(\ref{eq:average-x2}) is approximately equal to $\frac{1}{12} \phi_p^3$. 
We have not  expanded the cosine and sine factors  in  (\ref{eq:Pmumu3}), (\ref{eq:average-x3}) since $\phi$ 
or its deviations from $n \pi$  could become smaller than $\phi_p$ at certain moments of time. \\

According to (\ref{eq:average-x3}) the contribution to  the distance is proportional 
to~\footnote{
Though $\gamma_\pi$ and $E$ are related with each other  in pion decay 
we represent the both factors independently for possible use of the formula 
for neutrinos from other sources such as muon decay, {\it etc.}.
}
\be
\sigma \phi_p = 
1.25 \times 10^{-2} \text{cm }\left( \frac{ \Delta m^2 }{ 10^{-3} 
\text{eV}^2} \right) \left( \frac{ E }{ 1 \text{GeV} } \right)^{-1} 
\left( \frac{ \gamma_\pi }{ 10 } \right)^{- 2} 
 \left( \frac{ l_p }{ 100 \text{m} } \right)^2. 
\label{eq:parent}
\ee

Thus, in the limit $\Delta v \rightarrow 0$ the total distance traveled by neutrinos 
equals
$$
d(t, t_0) = v (t-t_0) + \left[\langle x(t) \rangle_{osc} - 
\langle x(t_0)\rangle_{osc} \right], 
$$
where the first term is the distance traveled by a 
massive neutrino. The second term takes the largest positive value 
$\approx  2 \times 10^{-3}$cm,  at around  $\phi \sim \frac{3 \pi}{2}$ and 
for the typical values of the parameters in the 
parentheses of eq. (\ref{eq:parent}). This term dominates over the first one.

\subsection{Estimation of effect of shift of the wave packets}
\label{sec:asymm-shape}

For realistic experimental setups $\Delta v t \ll \sigma <  l_{decay}$. 
The difference of group velocities of the mass eigenstates produces a relative shift 
of their wave packets and therefore an additional distortion  of the $\nu_\mu$ wave packet. Because of the shift, three different spatial parts of the $\nu_\mu$ wave packet appear (we assume that $v_1 > v_2$):

1). The front edge part: $(v_2 t + \sigma) \leq x \leq (v_1 t + \sigma)$.
In this part only the $\nu_1$ packet is present, therefore
$\psi_{\nu_\mu} \approx c^2 f(\sigma) =  c^2 f_0$ and $|\psi_{\nu_\mu}|^2 = c^4 f_0^2$. 
The length of this part equals to $\Delta v t$ and 
we can safely neglect the change of $f(x)$ within this interval. 

2). The overlapping part: $(v_1 t) \leq  x \leq (v_2 t + \sigma)$. 
Here both wave packets are non-zero and interfere between each other:
\be
|\psi_{\nu_\mu}|^2 \approx 
f(x)^2 \left|c^2 + s^2 \cos\Phi(x) 
\left(1 + \frac{\Delta v t \Gamma}{2(v - v_\pi)}\right)\right|^2 . 
\label{eq:cordelta}
\ee

3). The trailing edge part: $(v_2 t) \leq  x \leq (v_1 t)$. 
Here only the $\nu_2$ wave packet is present and $\psi_{\nu_\mu} \approx  s^2 f(0)$,  
so that $|\psi_{\nu_\mu}|^2 \approx  
s^4 f_0^2 e^{- \frac{\Gamma \sigma}{v - v_\pi}} $.

Apparently, oscillations do not affect contributions from 
the front and trailing edge parts.   
These parts produce additional asymmetry  of the $\nu_\mu$ wave packet 
which is proportional to 
$$
f_0^2 \left( c^4 - s^4  e^{- \frac{\Gamma \sigma}{v - v_\pi}} \right) \Delta v t 
= \left( c^4 - s^4  e^{- \frac{\Gamma \sigma}{v - v_\pi}} \right) 
\left(\frac{y}{1 - e^{-y}} \right) \frac{\Delta v t}{\sigma}. 
$$
Consequently, the contribution to the distance of  $\nu_{\mu}$ propagation 
$d_{shift-a} \sim \Delta v t$. As we will show later  the asymmetry produced by the  overlapping region  
(which depends on oscillations) is even smaller than  $\frac{\Delta v t}{\sigma}$. 
Therefore in the first approximation we can neglect dependence 
of the oscillation   effect inside the wave packet size 
on distance and take the phase $\Phi$ to be constant.

Let us introduce the dimensionless parameter 
\be
\epsilon \equiv \frac{\Delta v t \Gamma }{2 v_\pi} 
= \frac{\Delta v t }{2} \frac{1}{l_{decay}}
\ee
which is the half-shift of the wave packets in units of the decay length.  
Its value can be estimated as
\be 
\epsilon \approx 
\frac{\Delta v t }{2} \frac{\Gamma_0}{v_\pi \gamma_\pi} 
\approx  
3.20 \times 10^{-18} \left( \frac{ \Delta m^2 }{ 10^{-3} 
\text{eV}^2} \right) \left( \frac{ E }{ 1 \text{GeV} } \right)^{- 2} 
\left( \frac{ L }{ 10^{3} \text{km} } \right) 
\left( \frac{ \gamma_\pi }{ 10 } \right)^{- 1}.
\label{epsilon-value}
\ee
where $L = c t$ stands for baseline distance. 

When the oscillation effect in the coordinate space within the wave packet size is neglected 
we have $J = \cos \phi \int dx ~x ~f_1(x - v_1 t) f_2(x - v_2 t)$  and 
$I = \cos \phi \int dx ~f_1(x - v_1 t) f_2(x - v_2 t)$.  
Then straightforward computations give 
\begin{eqnarray}
I &=& \cos \phi 
\frac{e^{-\epsilon} - e^{\epsilon - y}}{1 - e^{-y}}, 
\nonumber \\
J &=& \cos \phi 
\frac{1}{1 - e^{-y}}\left[  
(e^{-\epsilon} - e^{\epsilon - y})\left( vt - \frac{\sigma}{y} \right) 
- \frac{\Delta vt }{2} (e^{-\epsilon} + e^{\epsilon - y})
+ \sigma e^{-\epsilon}\right]. 
\label{I-J-result}
\end{eqnarray}
For the average  distance traveled by the mass eigenstates we have  
$\bar{x}$ given in (\ref{eq:bar-x}) with $v = 0.5(v_1 + v_2)$.

Using (\ref{I-J-result}) we obtain correction to the distance of $\nu_\mu$ propagation 
\begin{eqnarray}
\langle x (t) \rangle_{shift-a} = \cos \phi  
~\frac{\sin^2 2\theta  }{2 P_{\mu\mu}} 
\left( \frac{\Delta v t }{2} \right) F(y,\epsilon),   
\label{eq:cor-shi}
\end{eqnarray}
where 
\begin{eqnarray}
F(y,\epsilon) \equiv  \frac{1}{(1 - e^{-y})^2}  
 \left[ y e^{-y}  \frac{e^{-\epsilon} - e^{\epsilon}}{\epsilon}
- (1 - e^{-y})(e^{-\epsilon} + e^{\epsilon - y}) \right], 
\label{eq:f-funct}
\end{eqnarray}
and  the  probability  $P_{\mu\mu}$ is given in (\ref{eq:Pmumu1}) with $I$
from (\ref{I-J-result}).

As can be seen from (\ref{epsilon-value}), 
$\epsilon$ is negligibly small 
for typical values of experimental 
parameters. In the limit of small $\epsilon$ we have 
\begin{eqnarray}
F(y)  =  \frac{1}{(1 - e^{-y})^2} \left[ 2y e^{-y}  -1  + e^{-2y} 
\right].  
\label{eq:f-funct2}
\end{eqnarray}
Also in most of the experimental settings $y \sim 1$, and consequently  
$F(y,\epsilon) \sim \mathcal{O}(1)$.  In particular,  $F(1) = 0.322$. 
Then according to (\ref{eq:cor-shi}) $\vert d(t) \vert 
\approx \langle x (t) \rangle_{shift-a} 
\sim \Delta v t$, assuming $\cos \phi \sim \mathcal{O} (1)$ 
and $P_{\mu\mu} \sim \mathcal{O} (1)$. 
Notice that $\langle x (t_0) \rangle$ term in $d(t)$ can be ignored  
for baselines $L  > 100$ km, since $l_p \ll L$. 
Numerically,   
\be
\Delta v t \approx  
\frac{ \Delta m^2 }{ 2 E^2} L = 0.5 \times 10^{-13} 
\left( \frac{ \Delta m^2 }{ 10^{-3} \text{eV}^2} \right) 
\left( \frac{ E }{ 1 \text{GeV} } \right)^{- 2} 
\left( \frac{ L }{ 10^{3} \text{km} } \right) \text{ cm }. 
\label{Delta-v-t}
\ee
Therefore, contribution to $d(t)$ due to velocity difference 
of mass eigenstates  is extremely small. 

For small $y$ we obtain $F \approx - y/3$.  In the limit of $y \rightarrow 0$, 
which correspond to long lifetime compared to the travel time inside decay tunnel, 
the function $F$ and therefore the distance vanish: 
$\langle x(t) \rangle_{shift} = 0$.  
This limit corresponds to symmetric (box-like) wave packet. 
Therefore, the contribution to the distance (\ref{eq:cor-shi}) described here originates from  
asymmetric profile of the shape factors. 

If $y \rightarrow \infty$, we obtain $F \rightarrow -1$. 
Let us underline  that $F < 0$, and therefore the correction is negative (reduces 
the effective velocity). It can be positive if formally $\Gamma < 0$, that is,  
when the amplitude  of the shape factor is a decreasing function of $x$, or 
equivalently, increases with time.  

Let us now estimate corrections to the distance due to the overlapping part 
by taking into account the coordinate-dependent oscillation effect. 
The correction due to $\Delta v t$ in (\ref{eq:cordelta}) does not depend on 
$x$,  and therefore can be ``absorbed'' in redefinition of the 
mixing parameters: 
$$
|\psi_{\nu_\mu}|^2 \approx 
(1 + r s^2)^2 f(x)^2 \left|c_1^2 + s_1^2 \cos\Phi(x) \right|^2,  
$$
where 
$$
c_1^2 = \frac{c^2}{1 + r s^2}, ~~~ s^2_1 = \frac{s^2 (1 + r)}{1 + r s^2},   
$$
and 
$$
r \equiv \frac{\Delta v t \Gamma}{2(v - v_\pi)} \sim \frac{\Delta v t}{\sigma}. 
$$
According to (\ref{eq:cor-shi}) the resulting additional distance equals 
$$
\langle x \rangle_{shift-a} \propto 
\frac{(1 + r s^2)^2 s_1^2 c_1^2}{P_{\mu\mu}} 
= (1 + r)s^2 c^2.  
$$
Therefore the correction to distance of propagation due to 
the shift is 
$$ 
\Delta \langle x \rangle_{shift-a}  = r \langle x \rangle_{shift-a}^0    
\sim \Delta v t \frac{\langle x \rangle_{shift-a}^0}{\sigma} \ll \Delta v t,
$$
where $\langle x \rangle_{shift-a}^0$ is the shift for constant 
survival probability. 

\section{Neutrino velocity in various experimental settings}
\label{sec:estimate}

\subsection{Distances for existing experimental  setups}

Here we present numerical estimates of the correction terms in $d(t)$ 
ignoring the length of pion decay tunnel compared to baseline distance, so that 
$d(t) \simeq \langle x (t) \rangle$. 
According to  (\ref{eq:decomposition}) 
the corrections to normal distance traveled by massive neutrinos 
consists of three different terms.  
It may be instructive to compare behavior of these terms 
in the limit of small phases $\phi \ll 1$ and $\phi_p \ll 1$ when 
all the contributions become linear in the time interval: 
\bea
d_{osc}~ & = & - \frac{\sin^2 2 \theta}{2 P_{\mu \mu}(t)}
\frac{\sigma \phi_p}{12} \frac{\Delta m^2}{2E}(t - t_0) 
= - \frac{\sin^2 2 \theta}{2 P_{\mu \mu}(t)}
\frac{\sigma  E \phi_p}{12} \Delta v (t - t_0),  
\nonumber\\ 
d_{shift-s} & = & \frac{\cos 2\theta}{2 P_{\mu \mu} (t)}
\Delta v (t - t_0),
\nonumber\\  
d_{shift-a} & = & \frac{\sin^2 2 \theta}{2P_{\mu \mu}(t)} F(y, \epsilon)
\Delta v (t - t_0).
\label{dees}
\eea
It follows immediately from these equations that $d_{osc} \gg d_{mass} \aprge  d_{shift}$. 
The contribution $d_{osc}$ is strongly enhanced by factor 
$\sigma E$  in comparison to the other contributions. 
%
%

All the terms in (\ref{eq:decomposition}) quickly decrease with energy: 
$$
d_{osc} \propto \frac{1}{E^4}, ~~ 
d_{mass} \propto \frac{1}{E^2}, ~~ 
d_{shift-s} \propto \frac{1}{E^2},~~
d_{shift-a} \propto \frac{1}{E^5}. 
$$
The contribution $d_{shift}$  vanishes for maximal mixing. 
Other contributions depend on $\theta$ rather weakly, or are independent of $\theta$.

In Table~\ref{v-correction} we present numerical estimate 
of the correction terms  $d(t)_i \approx \langle x \rangle_i$ 
for the existing experiments. We take  $\Delta m^2 = 2.5 \times 10^{-3} \text{eV}^2$, and $\sin^2 2\theta = 0.97$, which corresponds 
to the largest departure from the maximal mixing allowed by the Super-Kamiokande atmospheric neutrino data at 90\% CL \cite{SK-atom-I-III}.\footnote{
This is to avoid the maximal value of $\theta$ which leads to vanishing $d_{shift}$.
}
We use eqs. (\ref{eq:osc-corr}) and (\ref{eq:expeff})
to compute  $d_{osc}$. To find $d_{mass}$ we have assumed that 
$m_2^2 + m_1^2 \approx \Delta m^2 = 2.5 \times 10^{-3} \text{eV}^2$ 
which corresponds to the strong mass hierarchy.  
$d_{shift-s}$ has been computed according to eq. (\ref{eq:shift-s}).  
Notice that $d_{shift-s} = |d_{mass}|\cos 2\theta /P_{\mu\mu}$. 
As follows from (\ref{eq:cor-shi}) for the exponential (asymmetric) wave packet 
$$
d_{shift-a} = d_{shift-s} \frac{F(y)\sin^2 2\theta \cos\phi}{2 \cos 2\theta},  
$$
and therefore $d_{shift-a} \aprle d_{shift-s}$ unless $\cos 2\theta$ is very small. 

In Table~\ref{v-correction} we also show the size of the neutrino wave packet 
$\sigma$ which gives the absolute upper bound on additional contribution to the
distance of propagation, the parameter $y$ and the sine of oscillation phase.  
The phase $\phi_p$ acquired  by neutrinos in the wave packet 
formation region equals  $\phi_p = 4 \times 10^{-4}$  for $E = 17$ GeV in 
OPERA experiment and 
$\phi_p \sim 8.2 \times 10^{-4}$ in all the other setups with $y \sim 1$. 
The contributions have been computed for the average energy and for a representative value of $E$ at low energies in the spectrum in each experiment.  
We used the following baselines and decay tunnel lengths:  
OPERA: $L=730$ km and  $l_p = 1095$ m, MINOS: 
$L=735$ km, and  $l_p = 715$ m,   T2K:  $L=295$ km and $l_p = 118$ m. 

\begin{table}
\caption[aaa]{
The values of $y = l_{form}/l_{decay}$, sine of the oscillation phase, $\phi$, 
the wave packet length, $\sigma$, as well as the  contributions to the distance  
of $\nu_\mu$ propagation from the mass terms, $d_{mass}$, from the relative 
shift of the wave packets, $d_{shift-s}$,  and  from oscillations, $d_{osc}$.  
We use  $\Delta m^2 = 2.5 \times 10^{-3} \text{eV}^2$, and 
$\sin^2 2\theta = 0.97$. 
}
\vglue 0.5cm
\begin{tabular}{c|ccccccc}
\hline
\hline
Experiment   &\ \ $E$ (GeV) \
             &\ \ $y$ \  
	     &\ \ $\sin \phi$ \ 
             &\ \ $\sigma$ (cm) \
             &\ \ $- d_{mass}$ (cm) \
             &\ \ $d_{shift-s}$ (cm)  \
             &\ \ $d_{osc}$ (cm) \\
             \hline
OPERA            & \ \ 17
		 & \ \ 0.491
		 & \ \ 0.270
		 & \ \  0.671
                 & \ \  $1.58 \times 10^{- 16}$
                 & \ \  $2.80 \times 10^{-17}$
                 &\ \  $- 1.0 \times 10^{-5}$ \\
\hline
OPERA            & \ \ 1
                 & \ \ 1
		 & \ \  -0.99
		 &\ \ 23.3
                 &\ \ $4.56 \times 10^{-14}$
                 &\ \  $1.65 \times 10^{-14}$
                 &\ \ $+ 1.6 \times 10^{-3}$  \\
\hline
MINOS            & \ \ 3 
                 & \ \ 1
		 & \ \ 1.00
                 &\ \ 7.7
                 &\ \ $5.10 \times 10^{- 15}$
                 &\ \ $1.70 \times 10^{-15}$
                 &\ \  $- 4.6 \times 10^{-4}$ \\
\hline
MINOS            & \ \ 1 
                 & \ \ 1
		 & \ \ -0.995
                 &\ \ 23.3
                 &\ \ $4.59 \times 10^{-14}$
                 &\ \ $1.62 \times 10^{-14}$
                 &\ \  $+ 1.54 \times 10^{-3}$ \\
\hline
T2K              & \ \ 0.6
                 & \ \ 1
		 & \ \ 0.052
                 &\ \ 38.8
                 &\ \ $5.12  \times 10^{-14}$
                 &\ \ $2.94 \times 10^{-13}$
                 &\ \  $- 2.2 \times 10^{-3}$ \\
\hline
T2K              & \ \ 0.4 
                 & \ \ 1
		 & \ \ -0.997
                 &\ \ 58
                 &\ \ $1.14 \times 10^{-13}$
                 &\ \ $4.25 \times 10^{-14}$
                 &\ \  $+ 3.9 \times 10^{-3}$  \\
\hline
\hline
\end{tabular}
\label{v-correction}
\end{table}

For all the cases we find $d/\sigma \aprle 10 ^{-3}$. 
The superluminal motion is realized when $d_{osc} > |d_{mass}|$ and this condition can be satisfied for all existing setups. 
However, as follows from Table~\ref{v-correction}, the oscillation effects cannot explain the OPERA result in \cite{OPERA}: Indeed, we have shown that 

\begin{itemize}

\item
the additional distance of the  $\nu_\mu-$propagation is too small:   
It is restricted by size of the neutrino wave packet.  
Since the signal in OPERA is not suppressed, $P_{\mu \mu} \sim 1$, 
the effect cannot be related to oscillations into sterile neutrinos.  
Therefore, for the known $\Delta m^2$, the distance is further suppressed 
by the phase $\phi_p$ acquired by neutrinos in the production region.

\item 
the distance has strong dependence on neutrino energy $E$; $d$ increases as $E$ decreases. 

\end{itemize}

In MINOS at certain energies increase 
of propagation distance can be realized 
due to smallness of $P_{\mu \mu}$.  
In the case of $l_{decay} \ll l_p$ ($y \gg 1$) the effect can be enhanced 
if one includes in consideration the exponentially suppressed tails 
of the wave packets utilizing $l_p$  as the wave packet formation 
region instead of $l_{decay}$. (Of course, this requires  an extremely 
intense beam of neutrinos.)  From eq. (\ref{eq:expeff}) we obtain in the 
$y \rightarrow \infty$ limit 
\be
\langle x \rangle_{osc} \rightarrow 
- 2\pi \frac{\sin^2 2\theta}{2 P_{\mu \mu}} 
\frac{ l_{decay}^2 }{l_\nu}
 \left( \frac{v - v_\pi} {v_\pi}  \right) 
\left( \sin \phi + \phi_p \cos \phi \right)
\ee
So, in comparison to the case $y = 1$,  for $y \gg 1$ the distance increases 
by a factor $\eta_s(\infty)/\eta_s(1)  = 12.61$.
Therefore in all the cases in Table~\ref{v-correction}, where $y = 1$  
the distance would be at most $12.61$ times larger. In this way we 
obtain  $d_{osc} \sim 0.02$ cm for OPERA (1 GeV), MINOS (1 GeV) and T2K (0.4 GeV) 
instead of numbers in Table~\ref{v-correction}.
Even with such an  enhancement we have  $d_{osc}/\sigma < 10^{-3}$,  and 
observation of these distances looks practically impossible.

\subsection{Enhancement of contributions}

Let us consider a possibility to increase the additional 
distance  of $\nu_\mu$ propagation and to observe effective superluminal 
motion of $\nu_\mu$. With nanotechnology the accuracy of time measurement
would allow probing changes of distances of propagation as small as 1 m. 
Notice that one can evaluate the change of  time of $\nu_\mu$ propagation 
for a given baseline $L$ dividing the 
corrections to the distance of propagation  by $v$, e.g. 
$\Delta t_{osc} = - d_{osc}/v \approx - d_{osc}/c$.

In this discussion we will use the results for box-like 
shape factor which are simple and transparent,   
and at the same time are valid up to $\mathcal{O} (1)$ coefficients  
also for the exponential shape factors.  
The result for  the box-like wave packet 
(\ref{eq:average-x2}) can be rewritten in the following form 
\be
\langle x \rangle_{osc} =  
- \sigma  g(\phi_p) 
\frac{\sin^2 2\theta}{2 P_{\mu\mu}} 
\sin \left( \phi + \frac{\phi_p}{2} \right),  
\label{eq:average-new}
\ee  
where
\bea
g(\phi_p) \equiv 
\frac{2}{\phi_p^2} \left(\sin \frac{\phi_p}{2} - 
\frac{\phi_p}{2}\cos \frac{\phi_p}{2} \right),  
\label{eq:gfunc}
\eea
and  $P_{\mu\mu} \equiv P_{\mu\mu}(\phi + \frac{\phi_p}{2})$. 

According to (\ref{eq:average-new}) there are several ways to increase 
$d_{osc} \approx \langle x \rangle_{osc} $: 

\vspace{3mm}

1). Increase $\sigma = \sigma(\tau, l_{decay}, v_\pi, \theta_{\nu\pi}^0)$.  
For this one should lower the neutrino energy, use 
parent particles with long lifetimes, 
increase the size of decay region, and/or use off-axis neutrino beams.      
According to \cite{farzan}  
\be
\sigma_x = \frac{\tau_\pi^0}{\gamma_\pi(1 + v_\pi \cos \theta_{\nu\pi}^0)} , 
\ee
where $\theta_{\nu\pi}^0$ is the angle between a neutrino momentum in the 
pion rest frame and a momentum of pion in the laboratory frame. 
For $\theta_{\nu\pi}^0 = 0$ it reproduces our previous results. 
With increase of $\theta_{\nu\pi}^0$ the size of the wave packet increases. 
Simultaneously, the energy of neutrino decreases in such a way that 
$\sigma_x E$ is invariant~\cite{farzan}. E.g. for  $\theta_{\nu\pi}^0 = \pi/2$, 
$\sigma_x = 2 \sigma_x^0$ and energy becomes half.
Thus, for off-axis experiments the additional distance 
of propagation can be larger. 
For the backward emitted neutrinos, $\theta_{\nu\pi}^0 = \pi$, the size of the wave 
packet becomes the largest one: $\sigma_x = 2 \gamma_\pi  v \tau_\pi^0 = 
2  v \tau_\pi$. Then  $\sigma_x = 2 l_p$, if $l_{decay} > l_p$. 
The energy becomes $4\gamma_\pi^2$ times smaller, thus for 
$E_{\pi} = 1.4$ GeV it will be about 2 MeV. 

2). Increase of the phase $\phi_p = l_p(l_p, l_{decay}, l_\nu)$ acquired over the neutrino production region.  
This is possible with increase of $\Delta m^2$, decrease of the neutrino energy, 
{\it etc.}. 

3). Selection of certain value of the oscillation phase,  $\phi(L, l_\nu)$,  
by the selecting particular values of the baseline or/and neutrino energy. 
By varying the oscillation phase $\phi$ we find that maximal  value of the distance is given by 
\be
\langle x(t) \rangle_{osc}^{max} =  
- S~ \sigma ~ g(\phi_p) \frac{\sin^2 2\theta}{2  
\sqrt{(c^4 + s^4)^2 - 
\left(\sin^2 2\theta \frac{\sin \phi_p/2}{\phi_p} \right)^2}},   
\label{eq:xmax2}
\ee 
where $S \equiv sign[\sin(\phi + \phi_p/2) ]$  
and it is achieved at $\phi$ determined from the equation
\be 
\cos \left(\phi + \frac{\phi_p}{2} \right) = 
- \frac{\sin^2 2 \theta}{c^4 + s^4} 
\left( \frac{\sin\frac{\phi_p}{2}}{\phi_p} \right). 
\label{eq:cond}
\ee

4). Select baseline and/or neutrino energy so that 
$P_{\mu\mu}$ is small. 

\vspace{3mm}

Here we consider two possibilities to increase $d_{osc} \approx \langle x(t) \rangle_{osc}$. 
(i) Decrease of the survival probability $P_{\mu\mu}$ which  
implies strong suppression signal.  
(ii) Increase of  $\sigma$, $g(\phi_p)$ and $\sin \phi$ without 
substantial decrease of $P_{\mu\mu}$ and therefore the signal. 
Let us explore them in order. 

\vspace{3mm}

1).  Small $P_{\mu\mu}$ can be achieved by selecting 
certain values of $\phi(L, E)$. We consider the case of small $\phi_p$  
in which  
$$
g(\phi_p) = \frac{\phi_p}{12}. 
$$
Notice that  according to (\ref{eq:Pmumu2}) 
the  minimum of $P_{\mu\mu}$ is obtained at $\phi = \pi - \phi_p/2$ and equals  
$$
P_{\mu \mu}^{min} = \cos^2 2\theta + \sin^2 2\theta \frac{\phi_p^2}{48}.
$$
However, for this value of $\phi$  the distance 
$d_{osc}$ vanishes to the second order in $\phi_p$.  
On the other hand, for $\phi = \pi$ we obtain 
$$
P_{\mu \mu}^{min} = \cos^2 2\theta + \sin^2 2\theta \frac{\phi_p^2}{12},
$$
and the distance equals 
\be
d_{osc} = \frac{\sin^2 2\theta}{\cos^2 2\theta + 
\sin^2 2\theta \frac{\phi_p^2}{12}} \frac{\sigma \phi_p^2}{48}.
\ee
In the case of maximal mixing this equation gives $d_{osc} = \sigma/4$.  

In the limit $\phi_p \rightarrow 0$ we obtain from (\ref{eq:xmax2}) 
\be 
d_{osc}^{max} = \frac{\sigma \phi_p}{24}
\frac{\sin^2 2 \theta}{\sqrt{\cos^2 2\theta + \frac{\phi_p^2}{48} \sin^4 2\theta}}
\label{eq:maxxx}
\ee
which corresponds to 
$$
\cos \phi \approx -  \frac{2s^2 c^2}{c^4 + s^4}. 
$$ 
For maximal mixing we find from (\ref{eq:maxxx}) 
$$
d_{osc}^{max} = \frac{\sigma}{2 \sqrt{3}},   
$$
which can be considered as the maximal possible additional 
distance of propagation  due to oscillations. 
This, however, corresponds to the very small survival probability,  
$P_{\mu\mu} = \frac{\phi_p^2}{24} \sin^2 2\theta$.

For large $\phi_p$ and $\phi = \pi$ we obtain   
$$ 
d_{osc} = \frac{\sigma}{2} 
\frac{\sin^2 2 \theta}{\cos^2 2\theta + 
0.5 \sin^2 2\theta \left(1 -  \frac{\sin\phi_p}{ \phi_p}\right)} 
\left[ \frac{1 - \cos \phi_p}{ \phi_p^2} 
- \frac{\sin \phi_p}{2\phi_p} \right], 
$$
and therefore  $d_{osc} \aprle  \sigma/2$. 

2). Let us maximize other factors in 
(\ref{eq:average-new})  for $P_{\mu\mu} = \mathcal{O} (1)$ . 
The function $g(\phi_p)$ (\ref{eq:gfunc}) reaches maximum 
$$
g(\phi_p)  = \frac{1}{4} \sin \frac{\phi_p}{2} \approx 0.2 
$$
at $\phi_p$ determined from the condition 
$$
\tan (\phi_p/2) = \frac{\phi_p/2}{1 - \frac{(\phi_p/2)^2}{2}}. 
$$
The smallest value of the angle which satisfies this equation 
is $\phi_p \approx \pi + \epsilon$, where $\epsilon > 0$. (Other values give local extrema.) 
With further increase of  $\phi_p$ (above $\pi$) the function $g$ decreases 
as $1/\phi_p$, and for small $\phi_p$  it has  linear dependence: 
$g \propto \phi_p$.

The oscillatory factor in  (\ref{eq:average-new}) gives 
maximal distance of $\nu_\mu$ propagation  
when $\phi + \phi_p/2  = 3\pi/2$ so that  $\sin(\phi + \phi_p/2) = -1$. 
This, in turn,  can be achieved by selecting the baseline or/and neutrino energy. 
For  $\phi_p \approx \pi$  which corresponds to maximal value of $g$, 
one needs to  have $\phi = \pi + 2\pi k$ ($k$ is an integer). 
For these values of phase $\phi$ the probability is equal to the average probability: 
$P_{\mu\mu} = c^4 + s^4$. Consequently,  in the case of  maximal mixing we obtain 
\be
\langle x(t) \rangle_{osc} =  
\sigma  g(\pi) \approx 0.2 \sigma.  
\label{eq:zzz}
\ee  

For fixed value of $\phi_p$  the contribution $\langle x(t) \rangle_{osc}$ 
can be slightly larger if $\phi + \phi_p/2$ deviates from  $3\pi/2$. 
For $\phi_p = \pi$ and maximal mixing the equation (\ref{eq:xmax2})  gives 
$\langle x(t) \rangle_{osc} =  0.26 \sigma$. At the same time 
the probability becomes smaller than before: $P_{\mu \mu} \approx 0.3$.  
In essence, what we have here is an enhancement of the additional distance due to decrease of probability. 

Thus, the only way to obtain significant additional distance of $\nu_\mu$ 
propagation, which amounts to a significant fraction of the length of the 
wave packet size  
$\langle x(t) \rangle_{osc}/\sigma \sim \mathcal{O} (1)$   
without strong suppression of signal, is to increase the phase acquired 
by neutrinos over the production region,  
so that $\phi_p = \mathcal{O} (1)$.  

Let us consider possible setups  which can realize $\phi_p = \mathcal{O} (1)$ 
and increase $\sigma$. Apparently the region of formation of neutrino wave packet 
is restricted by the length of decay tunnel. At most  
$l_p = (1 - 2)$ km and therefore according to (\ref{eq:phi-p-value})
we have two possibilities: 

1).  For known $\Delta m^2 = 2.5 \cdot 10^{-3}$ eV$^2$ one can  
take low energies: $E =$ a few MeV. For muon neutrinos this can be arranged 
with  slowly moving  muons which have two orders of magnitude longer lifetime. 
The detection should then be via the neutral current interactions. Alternatively, 
one can detect backward moving neutrinos from the pion decay.  

For electron neutrinos one can use not only muon decay but also nuclei 
decay  (beta beams). $\nu_e$ detection via the charged current interactions 
is of course easier. In this case, however, the effect is suppressed by the 
small 1-3 mixing: $\sin^2 2\theta_{13} < 0.1$. 
It allows one to achieve 
$\langle x(t) \rangle_{osc}/\sigma \sim  (0.01 - 0.1)$. 
At low energies the size of the wave packet can be large: 
$\sigma \sim (0.1 - 1) l_p$, and therefore, 
$\langle x(t) \rangle_{osc} \sim (1 - 10)$ m. 
However, realization of these setups will require to overcome a significant 
loss of signal due to 
selection of low energy part of the muon decay spectrum, or the backward-going neutrinos, and the small cross-sections, {\it etc.}. 

2). If sterile neutrinos exist with large mass 
splittings, $\Delta m^2 \sim 1$ eV$^2$,   
the phase  $\phi_p$  can be large at higher energies: $E \sim 1$ GeV. 
Here again the effect will be suppressed by  
small allowed mixing:  $\sin^2 2\theta_{14} < 0.1$. 
As a result, $\langle x(t) \rangle_{osc}/\sigma \sim  (0.01 - 0.02)$
can be achieved. 

For low energy pions the size of the neutrino wave packet 
and the neutrino formation region are 
determined by the decay length, and therefore   
\be
\phi_p = \frac{\Delta m^2}{2E} \gamma_\pi \tau_\pi^0 v_\pi = 
\frac{\Delta m^2}{2m_\pi} \frac{E_\pi}{E} \tau_\pi^0.  
\ee
The phase depends weakly on energy and 
turns out to be of order unity for $\Delta m^2 > 10$ eV$^2$.  
The length of the wave packet equals  
$\sigma = \tau_\pi^0/ 2\gamma_\pi$ and for $E_\pi = 0.5$ GeV 
we obtain $\sigma \sim 1$ m. 
This gives  $\langle x(t) \rangle_{osc} \sim (1 - 2)$ cm.

Larger distance can be achieved 
for neutrinos from the muon decay. In the GeV energy range 
$l_{decay} > l_p$.  So, according to (\ref{eq:packet-size}) 
$\sigma \sim (10 - 20)$ m, and 
$\langle x(t) \rangle_{osc} \sim (10 - 20)$ cm. 

To summarize, we have shown that the large additional distance of $\nu_\mu$ propagation, $d_{osc} \sim (0.1 - 10)$ m,  can be obtained in 
rather non-standard experimental setups with  
low energy accelerators of muons and long decay tunnels, 
with use of the electron neutrinos or 
in the presence of large $\Delta m^2$. 

\section{Oscillation in matter and neutrino velocities}
\label{sec:matter}

In long-baseline experiments 
neutrinos propagate in matter. In this 
connection let us discuss influence of the matter effect
on the neutrino  velocity ~\cite{MS}. 
Notice that in the absence of mixing the dispersion relation in 
matter reads  
$$
E = \sqrt{p^2 + m^2} + V, 
$$
where $V$ is  the matter potential which can be written at low energies as 
$V_0 \approx \beta \sqrt{2} G_F n$ with  $G_F$ and $n$ being  the Fermi 
constant and the electron number density in matter, respectively. 
The  constant $\beta$ depends on the neutrino flavor. 
Since at low energies the potential $V_0$ 
does not depend on momentum of neutrino,  
$dV_0/dp = 0$, the group velocity remains unchanged 
in matter: 
$$
\frac{d E }{dp } = \frac{p}{E}. 
$$

$V$ depends on energy due to the $W$ boson propagator. 
For the elastic scattering in  forward direction $q^2 = 0$, and therefore the energy 
dependence of $V$ appears when the $W$ exchange occurs in the s-channel. 
In usual media this is possible for 
$\bar{\nu}_e$  only when $\bar{\nu}_e$ annihilates with electrons. 
In this case 
$$
V =  V_0 \frac{m_W^2}{m_W^2 - s} 
\approx  V_0 \frac{m_W^2}{m_W^2 - 2 m_e p - m_e^2},   
$$
and hence 
$$ 
\frac{dV}{dp} = V_0 \frac{2m_e m_W^2}{(m_W^2 - s)^2}. 
$$
If $s \ll m_W^2$ we obtain the energy independent contribution 
to velocity: 
\be
\Delta v \approx \frac{2 m_e V_0}{m_W^2} = 
1.8 \times  10^{-29} 
\left(\frac{\rho}{3 \text{g/cm}^3}\right) 
\left(\frac{Y_e}{  0.5 }\right), 
\label{Dv-matter}
\ee
where $\rho$ is the matter density and $Y_e$ is the electron fraction.
This conclusion does not depend on whether $V$ is independent of 
$x$ or not. 
The contribution to the neutrino velocity in (\ref{Dv-matter}) 
is too small to affect any of our discussions, and therefore it can be ignored.  

Since neutrinos are mixed, the propagating degrees of freedom are 
neutrino eigenstates in matter. The energy  
of these states  (eigenstates of the Hamiltonian) are given by 
$$
E_{1,2}^m = p + \frac{m_1^2 + m_2^2}{4p} +  \frac{V_1 + V_2}{2} 
\pm  \frac{1}{2} \sqrt{\left(V - 2\frac{\Delta m^2}{4p} \cos 2\theta \right)^2 + 
4 \left(\frac{\Delta m^2}{4p} \right)^2 \sin^2 2\theta}. 
$$   
Differentiating by $p$ we obtain
\be
v_{1,2}^m \approx 1 - \frac{m_1^2 + m_2^2}{4p^2} \mp \frac{\Delta m^2}{4p^2}
\frac{1 - \cos 2\theta \frac{2Vp}{\Delta m^2} }{\sqrt{
\left(\cos 2\theta  - \frac{2Vp}{\Delta m^2}\right)^2 + \sin^2 2\theta}} . 
\label{eq:velocity}
\ee 
In the limit of zero potential or small energies  it reproduces the usual result. 
At the resonance point, where the denominator is minimal, we find  
\be
v_{1,2}^m \approx 1 - \frac{m_1^2 + m_2^2}{4p^2} \mp \frac{\Delta m^2}{4p^2} 
\sin 2\theta . 
\label{eq:velocity1}
\ee 
At very high energies or large matter potential the velocities equal  
\be
v_{1,2}^m \approx 1 - \frac{m_1^2 + m_2^2}{4p^2} \pm \frac{\Delta m^2}{4p^2} 
\cos 2\theta .
\label{eq:velocity2}
\ee 
Thus,  as follows from (\ref{eq:velocity}) the correction 
to the velocity due to mixing in matter alone 
{\em cannot} lead to superluminal motion. 

For the  $\nu_\mu -\nu_\tau$ mixings in the limit
of zero 1- 3 mixing (neglecting loop corrections) the difference of potentials is zero,  $V = 0$,  and the  situation is reduced to 
the vacuum  case  described in Sec. II - IV.  
With non-zero $\theta_{13}$ in the three neutrino case the eigenvalues acquire additional dependence on momentum related to 
the matter potentials. Furthermore, the mixing angle and energy 
splitting in matter should be taken into account.  
However, at $E > 6$ GeV, {\it i.e.} above resonance energy 
in the Earth, this dependence is weak, and it can be neglected in the first approximation.  

It is possible to extend these statements to the case of full three flavor neutrino mixing.
The energy eigenvalues in matter can be written as
$E_{i}^m = p + \frac{ \lambda_{i} }{ 2p }$ ($i = 1, 2, 3$),
where $\lambda_{i}$ are given by~\cite{KTY2}
\begin{eqnarray}
\lambda_1&=&\frac{1}{3}s-\frac{1}{3}
\sqrt{s^2-3t}\left[u+\sqrt{3(1-u^2)}\right],
\nonumber \\
\lambda_2&=&\frac{1}{3}s-\frac{1}{3}
\sqrt{s^2-3t}\left[u-\sqrt{3(1-u^2)}\right],
\nonumber \\
\lambda_3&=&\frac{1}{3}s+\frac{2}{3}
u\sqrt{s^2-3t},
\label{lambda}
\end{eqnarray}
with 
\begin{eqnarray}
&&\hspace{-0.8cm}s=\Delta_{21}+\Delta_{31}+a ,
\nonumber \\
&&\hspace{-0.8cm}t=\Delta_{21}\Delta_{31}
+a[\Delta_{21}(1-s_{12}^2 c_{13}^2)+\Delta_{31}(1-s_{13}^2)],
\nonumber \\
&&\hspace{-0.8cm}u=\cos \left[\frac{1}{3}\cos^{-1}
\left(\frac{2s^3-9st+27a\Delta_{21}\Delta_{31}c_{12}^2 c_{13}^2}
{2(s^2-3t)^{3/2}}\right)\right].
\label{stu}
\end{eqnarray}
In (\ref{stu}) we have used the notations 
$\Delta_{ij} \equiv m_i^2-m_j^2$ and $a \equiv Vp$.
Notice that $p$-dependence of $\lambda_{i}$ is only through $a$.
Therefore,
\bea
v_i - 1 
= - \frac{ \lambda_{i} }{ 2 p^2 } + \frac{ 1 }{ 2p } \frac{ d \lambda_{i} }{ d a } \frac{ d a }{ d p}
= - \frac{ \lambda_{i} }{ 2 p^2 } + \frac{ 1 }{ 2 p^2 } \frac{ d \lambda_{i} }{ d a } a
= - \frac{ \lambda_{i} }{ 2 p^2 }
\left[ 1 -  \frac{ d ( \log \lambda_{i} ) }{ d ( \log a) } \right]. 
\label{velocity-3nu}
\eea
According to (\ref{lambda})  $\lambda_{i}$ is a monotonically increasing
function of $a$, but with the growth rate slower than $a$. 
(This feature can be seen in the plot for $\lambda_{i}$ as a function of $a$ given, e.g., in \cite{Dighe:1999bi,Minakata:2000rx}.)
Therefore, $\frac{ d ( \log \lambda_{i} ) }{ d ( \log a) } < 1$.
Asymptotically, the largest eigenvalue $\lambda_{3}$ behaves as $a$,
approaching to the equality.
Thus, $v_i - 1 < 0$ which excludes the possibility of neutrino's superluminal velocity due to matter effect.

\section{Conclusion}

Several factors alter the shape of the wave packet of a muon neutrino, and consequently, influence the distance of $\nu_\mu$ propagation for a given time and, hence, the velocity of a neutrino:  
(i) the relative shift of the wave packets of the mass eigenstates,  
(ii) oscillations, (iii) absorption, (iv)  production. 
In this paper, we  focussed  on the first two factors which are mutually correlated: 
both are due to the mass squared difference, and therefore 
the separation of the wave packets  is always accompanied 
by oscillations and {\it vice versa}.

1. We have computed the distances of $\nu_\mu$ propagation  
in the presence of mixing and oscillations. The 
oscillations lead to distortion of the  
shape factor of the $\nu_\mu$ wave packet. This, in turn, changes the effective 
distance traveled by neutrinos 
and therefore the group velocity. This is essentially related to  the oscillation effect 
within the neutrino wave packet.  
The oscillatory pattern is squeezed and therefore the effect is enhanced 
in the  same way as the size of the neutrino wave 
packet shrinks in comparison to the pion decay length or size of 
decay tunnel.  

We find that the distance of the $\nu_\mu$ propagation  
is proportional to the length of the wave packet $\sigma_x$
and the oscillation phase $\phi_p$ acquired by neutrinos
along the decay path of the parent particles  
(pions, K-mesons ) where neutrino wave packet is 
formed $d_{osc} \propto \sigma \phi_p$.
Furthermore,  $d_{osc}$ has an oscillatory behavior with distance determined by 
the oscillation length.  For small distances, $L < l_\nu/2$,  
the oscillations reduce the group velocity. 
The distance $d_{osc}$ becomes positive for baselines $L = l_\nu/2  - l_\nu$, {\it etc.}. 
In  this range of baselines 
motion can be  effectively superluminal.  
The additional distance is restricted by the size of  neutrino wave packet: 
$d_{osc} < \sigma_{x\nu}$.  The additional distance strongly decreases with increase of energy: $d_{osc} \sim 1/E^4$.

2.  Distortion of the  $\nu_\mu-$ wave packet
is also produced  due  to relative shift  of the wave  
packets of the mass eigenstates even in the case 
when oscillation effect does not depend on coordinate.  This effect 
previously considered in the literature is proportional 
to $\Delta v t = \Delta m^2/2E^2 t$, 
and therefore  negligible in comparison with the oscillation effect.

3. For the OPERA setup with $\Delta m^2 = 2.5 \cdot 10^{-3}$ eV$^2$ and 
$E \sim 17$ GeV, we obtain $d_{osc} \approx - 10^{-5}$ cm and other 
contributions are vanishingly small. 
Therefore, the OPERA result \cite{OPERA}, which corresponds to the distance $\sim + 20$ m,  cannot be explained. 
For average energy $E = 17$ GeV the oscillations reduce the distance propagated 
by neutrinos and consequently the group velocity. 
Furthermore, the distance $d_{osc}$ rapidly decreases  
with increase of the neutrino energy. 

4. We estimated the additional  distances of $\nu_\mu$ propagation 
for different experimental setups. In particular, we 
find $d_{osc} =  (0.01 - 0.04)$ cm  for MINOS  and T2K.
Change of the time of $\nu_{\mu}$ propagation (for a given $L$) can be 
obtained as $\Delta t \approx - d_{osc}/c$.

5. Larger additional distance  can be obtained for neutrinos 
from decays of particles with longer lifetime: muons, nuclei (beta beams).  
The $d_{osc}$  becomes larger with decrease of neutrino energy and for 
large $\Delta m^2$, if such exists. It can be as large as several meters. 
However,  this requires rather non-standard experimental setups and 
extremely intense neutrino fluxes.  
 
6. Measurements of  additional  distances of the flavor neutrino propagation 
opens a way to determine  sizes of neutrino wave packets 
and test certain  quantum mechanical features of neutrino production. 
In any case,  the oscillation effect should be taken into account 
in analysis of future measurements of the neutrino velocities. 

\begin{acknowledgments} 

This work was supported by the FY2011 JSPS Invitation 
Fellowship Program for Research in Japan, S11029, 
and by KAKENHI, Grant-in-Aid for Scientific Research No. 23540315, 
Japan Society for the Promotion of Science. A.Y.S. is grateful to 
D. Hernandez and E. Kh.  Akhmedov for useful discussions.

\end{acknowledgments}

\appendix
\section{Maximal effect of shift of the eigenstates}

Here we clarify the effect of shift of the wave packets
 discussed in the literature. 
Following \cite{Mecozzi}, \cite{Berry}, \cite{Indumathi}  
we take $\Delta p = 0$,  so that the oscillation phase  
is the same along the wave packet ({\it i.e.} by itself oscillations 
do not produce distortion of the $\nu_\mu$ packet).
For simplicity we take the box-like shape factors for the  mass eigenstates.  

Let us consider evolution of neutrino state produced as $\nu_\mu$. 
According to (\ref{eq:mixing}) in the $\nu_\mu$ state 
the $\nu_1$  and $\nu_2$ packets have amplitudes $c$ and $s$ respectively. (Here we omit normalization factors which are irrelevant for the final result.) 
The amplitude of probability to find $\nu_\mu$ 
in the $\nu_1$ part of the state is $c^2$ and 
in the $\nu_2$ part:  $s^2$. Result of measurement 
of $\nu_\mu$ in the full evolved neutrino state is then determined by interference
of the $\nu_\mu$ parts of $\nu_1$ and $\nu_2$.
The interference pattern, in turn, is determined by the relative (oscillation)
phase of the two mass eigenstates, $\Phi$,  and is described by $\cos \Phi$.
The phase does not depend on $x$ being the same over whole  wave packet.

At the moment of  time $t$ after production the centers  
of the $\nu_1$ and $\nu_2$ packets have coordinates $x = v_1 t + \sigma/2$ and 
$x = v_2 t + \sigma/2$ correspondingly. 
The average value of  the two coordinates is $ \bar{x} = \bar{v} t = (v_1 + v_2) t/2 + 
\sigma/2$.
The relative shift of the packets equals $\Delta v t  = \Delta m^2 t/2E^2$.
Correspondingly,  the front edges of $\nu_1$ and $\nu_2$  are at $v_1 t + \sigma$ and $v_2 t + \sigma$.

Consider  the shape of the $\nu_\mu$ wave packet  
and  $|\psi_{\nu_\mu}|^2$.
As we marked in Sec. IVC, due to the relative shift of the wave packets 
of mass eigenstates there are three different spatial parts of the
$\nu_\mu$ wave packet ($v_1 > v_2$):

1. In  the front edge part,  
$(v_2 t + \sigma) \leq x \leq  (v_1 t + \sigma)$  
only $\nu_1$ packet is present, so that 
$\psi_{\nu_\mu} =  c^2$ and $|\psi_{\nu_\mu}|^2 = c^4$.
The size of this part is given by the shift $\Delta v t$.

2. In the overlapping part, 
$(v_1 t)  \leq x  \leq  (v_2 t + \sigma)$,  
both wave packets are nonzero and 
$|\psi_{\nu_\mu}|^2 = |c^2    + s^2 \cos \Phi|^2 $.

3. In the trailing edge part, 
$(v_2 t) \leq x \leq (v_1 t)$, 
only $\nu_2$ wave packet is present and $\psi_{\nu_\mu} =  s^2$,  
so that $|\psi_{\nu_\mu}|^2 = s^4$.

In the first approximation the distance of $\nu_\mu$ propagation  
is determined  by  the position of the ``center of mass'' of the wave packet 
squared. 
In the case of maximal mixing whole the picture is completely symmetric with
respect to $\bar{x}$ and therefore  $\langle x \rangle = \bar{x}$.
If mixing deviates from maximal one the shift of the packets leads to
asymmetric distortion of the  $\nu_\mu$ wave packet.
This in turn, leads to a shift 
of the  ``center of mass'' from $\bar{x}$: 
$$
\langle  x \rangle = \bar{x} + \delta_x. 
$$
For non-maximal mixing (with $c > s$) the forward edge 
is higher by $c^4 - s^4 = \cos 2 \theta$  than the trailing edge. This difference should be 
compensated by the shift of the center of  mass by amount $\delta_x$ in the overlapping region.  This compensation leads 
to the condition 
$$
\Delta v t \cos 2 \theta = 2 \delta_x |c^2    + s^2 \cos \Phi|^2. 
$$
Therefore 
$$
\delta_x = \frac{\cos 2 \theta \Delta v t}{2 |c^2 + s^2 \cos \Phi|^2}.  
$$
The maximal shift would correspond to $\cos \Phi = - 1$. 
It implies the destructive interference in the overlapping region,  when
$\Phi = \pi$, so that $|\psi_{\nu_\mu}|^2 = |c^2 -  s^2 |^2  =  
\cos^2 2\theta $, and therefore 
\be
\delta_x = \Delta v t \frac{\cos 2 \theta}{2 \cos^2 2 \theta}
= \Delta v t \frac{1}{2 \cos 2 \theta},  
\ee
which reproduces results in \cite{Mecozzi}, \cite{Berry}, \cite{Indumathi}.
Thus,  a superluminal motion here is a result of interplay of 
the coordinate-independent oscillations and the relative shift 
of wave packets due to different group velocities.
Apparently the  shift  $\delta_x$ is restricted by the size of the wave packet. 
In this case it is clear that $v$ has no physical meaning. It is the velocity of the 
``center of mass'': some effective point in the flat overlapping part of the shape factor. 
Neither a single body nor real structure in the shape factor (edges of different regions) is moving with $v > c$.


\begin{thebibliography}{99}


\bibitem{Fermilab}
  J.~Alspector, G.~R.~Kalbfleisch, N.~Baggett, E.~C.~Fowler, B.~C.~Barish, A.~Bodek, D.~Buchholz and F.~J.~Sciulli {\it et al.},
  Phys.\ Rev.\ Lett.\  {\bf 36} (1976) 837.
  
  G.~R.~Kalbfleisch, N.~Baggett, E.~C.~Fowler and J.~Alspector,
  Phys.\ Rev.\ Lett.\  {\bf 43}, 1361 (1979).

\bibitem{MINOS} 
P. Adamson {\it et al.} [MINOS Collaboration], Phys. Rev. {\bf D76} (2007) 072005. 

\bibitem{OPERA}
  T.~Adam {\it et al.}  [OPERA Collaboration],
  arXiv:1109.4897 [hep-ex].

\bibitem{ICARUS} 
  M.~Antonello {\it et al.}  [ICARUS Collaboration],
  arXiv:1203.3433 [hep-ex].

\bibitem{SN1987a}
 K.~S.~Hirata, T.~Kajita, M.~Koshiba, M.~Nakahata, Y.~Oyama, N.~Sato, A.~Suzuki and M.~Takita {\it et al.},
  Phys.\ Rev.\ D {\bf 38}, 448 (1988).
  
\bibitem{many} 
See references to \cite{OPERA}. 

\bibitem{Garrett}
C.~G.~B.~Garrett and D.~E.~McCumber, 
Phys.\ Rev.\  A {\bf 1}, (1970) 305.

\bibitem{Chu}
  S.~Chu and S.~Wong,
  Phys.\ Rev.\ Lett.\  {\bf 48} (1982) 738.

\bibitem{Wang}
Wang Yun-ping and Zhang Dian-lin, Phys.\ Rev.\  A {\bf 52}, (1995) 2597.

\bibitem{Sanchez} 
M. M. Sanchez-Lopez, {\it et al.}, Appl. Phys. Lett. {\bf 93} (2008) 074102. 

\bibitem{Brunner} 
N. Brunner, {\it et al.}, Phys. Rev. Lett. {\bf 93}, (2004) 203902. 

\bibitem{Mecozzi}
  A.~Mecozzi and M.~Bellini,
  arXiv:1110.1253 [hep-ph].

\bibitem{Morris}
T.~R.~Morris,
  J.\ Phys.\ G G {\bf 39}, 045010 (2012)
  [arXiv:1110.2463 [hep-ph]].

\bibitem{Berry}
  M.~V.~Berry, N.~Brunner, S.~Popescu and P.~Shukla,
  J.\ Phys.\ A A {\bf 44} (2011) 492001
  [arXiv:1110.2832 [hep-ph]].

\bibitem{Indumathi}
 D.~Indumathi, R.~K.~Kaul, M.~V.~N.~Murthy and G.~Rajasekaran,
  Phys.\ Lett.\ B {\bf 709}, 413 (2012)
  [arXiv:1110.5453 [hep-ph]].

\bibitem{Tanimura}
  S.~Tanimura,
  arXiv:1110.1790 [hep-ph].
  
\bibitem{AHS}
E.~Akhmedov, D.~Hernandez and A.~Smirnov,
  JHEP {\bf 1204}, 052 (2012)
  [arXiv:1201.4128 [hep-ph]].

\bibitem{farzan}
Y.~Farzan and A.~Y.~Smirnov,
  Nucl.\ Phys.\ B {\bf 805} (2008) 356
  [arXiv:0803.0495 [hep-ph]].

\bibitem{SK-atom-I-III}
  R.~Wendell {\it et al.}  [Super-Kamiokande Collaboration],
  Phys.\ Rev.\ D {\bf 81} (2010) 092004
  [arXiv:1002.3471 [hep-ex]].

\bibitem{MS}
  S.~P.~Mikheyev and A.~Y.~Smirnov,
  Prog.\ Part.\ Nucl.\ Phys.\  {\bf 23} (1989) 41.
  S.~P.~Mikheev and A.~Y.~Smirnov,
  Sov.\ Phys.\ Usp.\  {\bf 30} (1987) 759
  [Usp.\ Fiz.\ Nauk {\bf 153} (1987) 3].

\bibitem{KTY2}
  H.~W.~Zaglauer and K.~H.~Schwarzer,
  Z.\ Phys.\ C {\bf 40} (1988) 273.

\bibitem{Dighe:1999bi}
  A.~S.~Dighe and A.~Y.~Smirnov,
  Phys.\ Rev.\ D {\bf 62} (2000) 033007
  [hep-ph/9907423].

\bibitem{Minakata:2000rx}
  H.~Minakata and H.~Nunokawa,
  Phys.\ Lett.\ B {\bf 504} (2001) 301
  [hep-ph/0010240].



\end{thebibliography}
\end{document}